\documentstyle[12pt,epsfig]{article}
 1
\def\citenum#1{{\def\@cite##1##2{##1}\cite{#1}}}
\def\citea#1{\@cite{#1}{}}
\def\beq{\begin{equation}}
\def\eeq{\end{equation}}
\def\bea{\begin{eqnarray}}
\def\eea{\end{eqnarray}}

\def\eq#1{{Eq.~(\ref{#1})}}

\begin{document}
\begin{titlepage}
\noindent
\begin{flushright}
December  1998 \\
TAUP 2542 - 98\\
DTP/98/102\\
\end{flushright}
\vspace{1cm}
\begin{center}
{\Large \bf  SURVIVAL PROBABILITY}\\[1.5ex]
{ \Large \bf  OF LARGE RAPIDITY GAPS}\\[4ex]

{\large E. ~LEVIN ${}^{a), 1)}$, A.D. ~MARTIN${}^{b),2)}$\,\,
and M.G. ~RYSKIN${}^{c), 3)}$}
 \footnotetext{$^{1)}$ Email: leving@post.tau.ac.il .}
\footnotetext{$^{2)}$ Email: A.D.Martin@durham.ac.il.}
\footnotetext{$^{3)}$ Email: RYSKIN@vxdesy.desy.de
}\\[4.5ex]
{\it a)\,\,\, School of Physics and Astronomy}\\
{\it Raymond and Beverly Sackler Faculty of Exact Science}\\
{\it Tel Aviv University, Tel Aviv, 69978, ISRAEL}\\[1.5ex]
{\it b)\,\,  Physics Department, University of Durham}\\
 {\it South Road, Durham DH1 3LE, UK}\\[1.5ex]
{\it c)\,\,\,  Theory Department, Petersburg Nuclear Physics Institute}\\
{\it 188350, Gatchina, St. Petersburg, RUSSIA}\\[3.5ex]
\end{center}
~\,\,\,
\vspace{2cm}

{\large \bf Abstract:}
We summarize the discussions on the value of the survival probabilities of
the large rapidity gap ( LRG ) processes at Durham WS'98.

\end{titlepage}

{ \bf 1.} When we  calculate the cross sections with the LRG such as  the 
diffractive (double
diffractive (DD)) processes or two jets production with the LRG between
them,  we have to estimate the probability that
rapidity gap corresponding to the Pomeron exchange will not be filled
by secondaries from the scattering of a spectator partons or
 from the decay of bremsstrahlung gluons. This survival
probability $<S^2>\,\,<\,\,1$ depends on the kinematics of each specific
process
and may violate the factorization (even for the cases where the
factorization properties of diffractive amplitudes are satisfied at
the initial stage).\\

{ \bf 2.} To demonstrate the role of bremsstrahlung emission let us
consider 
the exclusive Higgs boson DD production
\begin{equation}
pp\;\to\; p\; + [ LRG ]\,+ \; H\; +\;[ LRG ]\,+\,   p
\label{a1}
\end{equation}
where the heavy boson $H$ is bounded by two 
rapidity gaps \cite{KMRh}.

We start from the simplest Low-Nussinov (two gluon exchange) model for the 
Pomeron. The Higgs boson is produced in the central rapidity region by the 
gluon-gluon fusion ($gg\to H$) whilst a second t-channel gluon is needed to 
screen the colour flow. The amplitude is proportional to the integral over the 
gluon transverse momentum $q_t$
\begin{equation}
A\;\sim\;\int_{q_0}\frac{dq^2_t}{q^4_t}
\label{a2}
\end{equation}
The infrared cutoff $q_0$ provided by  confinement ($q_0\sim 1/R$) is of 
the order of the inverse proton radius $R$. In this picture the two t-channel 
gluons form a colour dipole with  size $r\sim 1/q_t$. When two such dipoles 
annihilate into the Higgs boson they normally emit bremsstrahlung gluons with 
$k_t>q_t$ (but with $k_t$ much less than Higgs mass $M_H$).

 In the Double Log approximation the mean number of emitted gluons is
 \begin{equation}
n\;\simeq\;\frac{N_c\alpha_s}{2\pi}\ln^2\frac{M^2_H}{4q^2_t}\;\;\;\;\; ,
\label{a3}
\end{equation}
and the probability amplitude not to observe these extra gluons is
$$<S^2 >_{ bremsstrahlung}\,\,=\,\,e^{-n/2}\,\,.$$

We have to include this factor $< S^2 >$ in the integral (\ref{a2}). Now
the 
low $q_t$ contribution is suppressed and the integral has a saddle point 
at rather large $q^2_t=q^2_0$. In other words the Sudakov form factor $< 
S^2 >_{ bremsstrahlung}$ plays a role of the sirvival probability 
and selects 
the small size components of the Pomeron wave function (that is 
the configurations 
which do not emit too many gluons). Of course the cross section becomes smaller
 for such a small size Pomeron; e.g. for $M_H\sim 200$ GeV  reaction 
(\ref{a1}) is 
suppressed by more than 1000  (!) by the pure perturbative ($q^2_t\sim q^2_0
\sim 15$ GeV$^2$) effects.

For more a complicated process the effective form factor $< S^2 >$ depends
on
 the specific 
kinematics (see \cite{KMRd} for the case of high $E_T$ dijet DD production) 
and clearly violates the factorization property of Pomeron exchange 
amplitudes.\\

{ \bf 3.}  The probability $< S^2 >_{spectators} $ to avoid/(not to
observe)
the rescattering of 
spectator partons may be estimated by extending "Pumplin" bound \cite{P}:
$\sigma^D\le\sigma_{tot}/2$ (here $\sigma^D=\sigma_{el}+\sigma^{SD}+\sigma^{DD}$
 is the sum of all diffractive cross sections -- elastic, single and double 
 diffractive dissociation).
 
 Using the eikonal model for parton-parton rescattering, and averaging over the 
 impact parameter $b_t$, one obtains the expression \cite{RR}
\begin{equation}
 < S^2 >_{spectators}\,\; =\,\; \left(
1-2\frac{2\sigma^D}{\sigma_{tot}}\right)^2
\label{a4}
\end{equation}
which can be used for phenomenological estimates of the survival 
probability $< S^2 >_{spectators}$.

However, some times the correlations in the $b_t$-plane becomes important
both for the value of the survival probability and its energy dependence
\cite{GLM1} \cite{GLM2}. To illustrate this point we take the simple
Eikonal model in which the survival probability $< S^2 >_{spectators}$ has
a very  transparent form \cite{BJ} \cite{GLM1}:
\beq \label{EK1}
< S^2 >_{spectators}\,\,  =\,\, \frac{ \int d^{2}b \Gamma_{H}(b) P(s,b)}
{ \int d^{2}b \Gamma_{H}(b)}\,\,,
\eeq
where the "hard" profile
\beq \label{EK2}
\Gamma_{H}(b) = \frac{1}{\pi R^{2}_{H}(s)}e^{-\frac{b^{2}}{R^{2}_{H}(s)}}
 \eeq
where $R_{H}$ denotes the radius of interactions in the "hard" scattering
process ( for example, two jet production with high transverse momenta and
LRG between them ) .  $P(s,b) = e^{- \Omega(s,b)}$ is the probability that
no inelastic  
interaction takes place at impact parameter $ b$.
Indeed,this meaning of $P(s,b)$ follows directly from  the unitarity
constraint which gives for the probability of all inelastic interaction
$G_{in} (s, b )$: 
\beq \label{EK3}
G_{in}(s,b) = 1- e^{- \Omega(s,b)}\,\,,
\eeq
where $ \Omega(s,b)$ is an arbitrary real function called opacity.

Assuming the simplest $b$-profile for opacity:
\beq \label{EK4}
\Omega(s,b)\,\,\,=\,\,\nu(s)\,e^{- \,\frac{b^2}{R^2_S(s)}}\,\,,
\eeq
with $R^2_S(s)\,\,=\,\,2 \,B_{el} (s)$   ( i.e. $ \frac{d \sigma}{dt} \sim
e^{B_{el}(s)t}$), one can see \cite{GLM1}\cite{GLM2} that the ratio
$R\,\,=\,\,\frac{\sigma_{el}}{\sigma_{tot}}$ depends only on $\nu(s)$,
namely:
\beq \label{EK5}
R\,\,=\,\,\frac{\sigma_{el}}{\sigma_{tot}}\,\,=\,\,1\,\,-
\,\,\frac{\ln(\nu)\,+\,C\,-\,Ei( - \nu )}{2\,[ \,\ln(\nu/2)\,+\,C\,-\,Ei(
- \frac{\nu}{2} )\,]}\,\,.
\eeq
\eq{EK5} allows us to find the value and energy dependence of parameter
$\nu(s)$ using the experimental data for ratio $R$.

Using \eq{EK5} we can calculate the survival probability due to possible
interaction of spectator partons ( see \eq{EK1} ):
\beq \label{EK6}
< S^2 >_{spectators}\,\,=\,\, \frac{a(s) \gamma[a(s),
\nu(s)]}{[\nu(s)]^{a(s)}}\,\,,
\eeq
 where $ Ei(x) = \int_{- \infty}^{x} \frac{e^{t}}{t} dt $
is integral exponent, $ \gamma(a,x) = \int_{0}^{x} z^{a-1}e^{-z}dz $ is
the incomplete gamma function,  C= 0.5773 is the Euler constant and 
$a(s)$ is defined by
\beq \label{EK7}
a(s) =\frac{R^{2}_{S}(s)}{R^{2}_{H}(s)}
\eeq
In Ref. \cite{GLM2} the value of $R_H$ has been evaluated ( $ R^2_H
\,=\,8\,GeV^{-2}$ ) using the
experimental data on doupble parton cross section and on the J/$\Psi$
production in DIS.
\begin{figure}[cont]
\epsfig{file=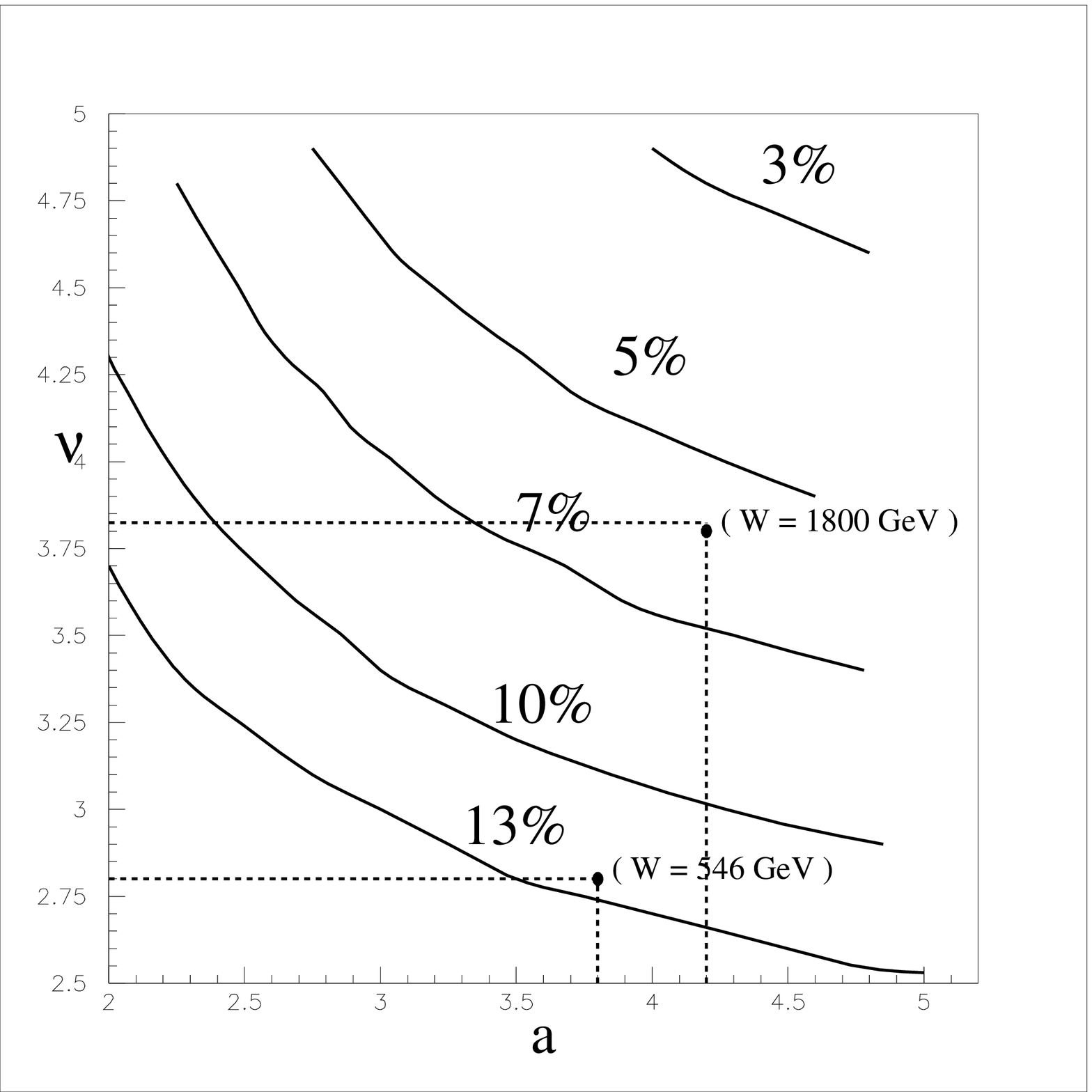,width=140mm}  
\caption{ Contour plot of survival probality $ <  S^2>_{spectators} $.}
\label{Fig.1}
 \end{figure}
Fig. 1 shows the result of numerical calculations \cite{GLM2} which
shows that this simple model can reproduce a small value of the survival
ptobability and its energy dependence, namely:
$$ \frac{< S^2>_{\sqrt{s}=630}}
{< S^2>_{\sqrt{s}=1800}} \;\;= \;\; 2.2\;\pm 0.2 \,\,,$$
 which is in a perfect agreement with the experimenal data \cite{D0}.

{\bf 4.} The resulting survival probability $< S^2 > \,\,=\,\,< S^2
(\Delta y  >_{ bremsstrahlung}\,\times\,< S^2 (s)>_{spectators}$. The
first factor depends mostly on the value of the LRG and can be estimateed
in the framework of pQCD, the second one depends mostly on the total
energy and could be evaluated only in nonperturbative QCD. It means,
practically, that we can have only models for its calculation. The
Eikonal model with Gaussian impact parameter dependence is one of many,
which has a obvious shortcoming since it takes into account only
interaction of the fastest spectators. The only way to avoid uncertainties
in calculations of $< S^2 (s)>_{spectators}$ which we see at the moment is
to measure the LRG processes in DIS \cite{ELLRG}. In DIS $ < S^2
(s)>_{spectators}$ is about 0.7 - 0.8 and can be calculated in pQCD
\cite{ELLRG} \cite{GLMSM}. For larger value of $Q^2$ $< S^2
(s)>_{spectators}\,\,\longrightarrow\,\,1$ and it makes  LRG processes in
DIS  is a good laboratory to measure $ < S^2 (\Delta
y)>_{bremsstrahlung}$.


\begin{thebibliography}{99} 
\bibitem{KMRh} V.A.Khoze, A.D.Martin, M.G.Ryskin, Phys.Lett. B401 (1997) 330.
\bibitem{KMRd} V.A.Khoze, A.D.Martin, M.G.Ryskin, Phys.Rev. D56 (1997) 5867.
\bibitem{P} J.Pumplin, Phys. Rev. D8 (1972) 2899. 
\bibitem{RR} A.Rostovtsev, M.G.Ryskin, Phys.Lett. B390 (1997) 375.
\bibitem{GLM1} E. ~Gotsman, E. ~Levin and U. ~Maor: Phys. Lett. B309 (
1993 ) 199.
\bibitem{GLM2} E. ~Gotsman, E. ~Levin and U. ~Maor: TAUP 2485, {\tt hep -
ph/9804404}, Phys. Lett. ( in press ).
\bibitem{BJ} J.D. ~Bjorken: Int. J. Mod. Phys. A7 ( 1992 ) 4189; Phys.
Rev. D47 ( 1993 ) 101.
\bibitem{D0} 
A. Brandt at  Workshop on "Interplay between Soft and Hard
interactions in DIS", Heidelberg Germany, September 1997.
\bibitem{ELLRG}
E. Levin:  Phys. Rev. D48 ( 1993 ) 2087.
\bibitem{GLMSM}
E. ~Gotsman, E. ~Levin and U. ~Maor: Nucl. Phys. B493 ( 1997 ) 354.

\end{thebibliography}
\end{document}